\documentclass[useAMS,usenatbib]{mn2e} 
\usepackage{psfig}

 \usepackage{times}

\def\kms{km ${\rm s}^{-1}$}

\def\ch2{$\chi^2$}

\def\kms {\hbox{${\rm km\ s}^{-1}$}}

\def\scm  {$\hbox{{\rm cm}}^{-2}$}    

\def\MOLH {\hbox{${\rm H}_2$}}  

\def\lapp{\ifmmode\stackrel{<}{_{\sim}}\else$\stackrel{<}{_{\sim}}$\fi}
\def\gapp{\ifmmode\stackrel{>}{_{\sim}}\else$\stackrel{>}{_{\sim}}$\fi}

\title[Molecular fraction limits in damped Lyman-alpha absorption systems]{Molecular fraction limits in damped Lyman-alpha absorption systems}
   \author[S. J. Curran et
    al.]{S. J. Curran$^{1}$\thanks{E-mail: sjc@phys.unsw.edu.au}, M. T. Murphy$^{2}$, Y. M. Pihlstr\"{o}m$^{3}$, J. K. Webb$^{1}$, \newauthor A. D. Bolatto$^{4}$  and G. C. Bower$^{4}$\\$^{1}$School of Physics, University of
    New South Wales, Sydney NSW 2052,
    Australia\\$^2$Institute of Astronomy, Madingley Road,
Cambridge CB3 0HA, UK\\$^{3}$National Radio Astronomy Observatory, Socorro, NM
    87801, USA\\$^{4}$Radio Astronomy Lab, 601 Campbell Hall, University of California, Berkeley, CA 94720, USA}

\begin{document}

\date{Accepted ---. Received ---; in original form ---}

\pagerange{\pageref{firstpage}--\pageref{lastpage}} \pubyear{2004}

\maketitle

\label{firstpage}

\begin{abstract}
We have used the Green Bank Telescope (GBT) and
Berkeley-Illinois-Maryland Association (BIMA) array to search for
redshifted millimetre absorption in a sample of damped Lyman-alpha
absorption systems (DLAs). This brings the number of published systems
searched from 18 to 30. In 17 cases we reach $3\sigma$ limits of
$\tau\leq0.1$, which is a significant improvement over the previous
searches and more than sufficient to detect the 4 known redshifted
millimetre absorbers ($\tau\gapp1$). While the CO rotational
(millimetre) column density limits obtained are weaker than the
electronic (optical) limits, they may provide useful limits below the
atmospheric cut-off for the Lyman and Werner \MOLH-bands in the UV
($z_{\rm abs}\lapp1.8$). Using a model for the DLA metallicity
evolution combined with assumed HCO$^+$/\MOLH ~and CO/\MOLH
~conversion ratios, we use the molecular column density limits to
calculate plausible \MOLH ~molecular fraction limits. Finally, we use these
results to discuss the feasibility of detecting rotational CO
transitions in DLAs with the next generation of large radio
telescopes.
\end{abstract}


\section{Introduction}\label{sec:intro}

Molecular absorption lines trace, and provide detailed physical and
chemical information about, the cold dense component of the
interstellar medium (ISM).  Although many detailed studies exist for
molecular clouds within our own Galaxy, only relatively recently has
detailed information emerged for molecular abundances at high
redshift, through absorption studies of redshifted UV molecular
hydrogen lines (e.g. \citealt{lps03,rbql03}) and millimetre-band
rotational lines from molecular tracers (e.g. \citealt{wc96a}).

Observations of a range of different molecular transitions in gas
clouds at high redshift would provide a wealth of information on star
formation activity in external galaxies, potentially viewed at epochs
when chemical abundances and environments were markedly different to
today.  Such information is invaluable for a detailed understanding of
galactic formation and evolution. Furthermore, the narrowness of
molecular lines reveals information about small-scale structure in the
ISM. Comparison of relative line strengths yields information about
the excitation mechanism, and in particular, probes the temperature of
the Cosmic Microwave Background (CMB), and hence the expected
($1+z$)--dependence (e.g. \citealt{wc96a}). Finally, comparisons of
the relative observed frequencies of millimetre molecular lines (with
each other, and/or with atomic transitions arising in the same cloud)
can be used to check on any possible variation in certain combinations
of the fundamental constants \citep{dwbf98,mwf+00}. This last point
was the prime motivation for our search for new redshifted millimetre
absorbers reported in this paper, although we use the upper limits
obtained to yield molecular fraction limits in low redshift DLAs,
which is not possible using the UV \MOLH ~lines, which fall below the
atmospheric cut-off of $z_{\rm abs}\lapp1.8$.

Currently, only 4 redshifted millimetre absorption systems are known
(see \citealt{wc99b} and references therein), of which the highest
redshift is 0.886 (PKS 1830--211). As a means of approaching a search
for new high redshift radio absorbers systematically, we produced a
catalogue of DLAs \citep{cwbc01}\footnote{A version of this catalogue
is kept updated on-line and is available from
http://www.phys.unsw.edu.au/$\sim$sjc/dla}, where large column
densities ($N_{\rm HI}\geq2\times10^{20}$ cm$^{-2}$) are known to
exist, and shortlisted those which are illuminated by radio-loud
quasars (i.e. those with a measured radio flux density $>0.1$
Jy). This yielded 60 DLAs and sub-DLAs occulting radio-loud
quasars. Of these, 37 have been searched for 21-cm absorption (see
\citealt{kc02,cmp+03}). Selecting those of 12-mm and 3-mm flux
densities $\gapp0.1$ Jy, gives 18 systems which have previously been
searched for millimetre absorption (\citealt{cmwp02}), this number now
being increased in total to 30 with the observations we present in
this paper.

\section{Observations and Data Reduction}\label{subsec:obs}

\subsection{12-mm observations}

In Table \ref{gbt}\footnote{Unless otherwise stated, B1950.0 names are
used throughout this paper.} we list all of the known DLAs and
sub-DLAs which are illuminated by background sources of $\gapp0.1$ Jy
at centimetre wavelengths and have molecular transitions which could
be detected in the GBT K-band.  In all cases this is both the HCN and
HCO$^+$ $J=0\rightarrow1$ rotational transitions\footnote{Due to the
high dipole moment of the density tracing molecules (i.e. HCN and
HCO$^+$), these will generally be close to the ground state causing
the opacity to rival that of the traditional molecular tracer CO which
will be excited to higher rotational levels by the CMB at these
redshifts.}. We chose to observe the latter transition since this
tends to be slightly stronger than HCN in the 4 known millimetre
absorption systems.
\begin{table}
\centering
\begin{minipage}{78mm}
\caption{All of the 22 GHz illuminated DLAs and sub-DLAs for which a strong
molecular rotational transition falls into the GBT K-band (range
18--26.5 GHz): Prior to the results of \protect\citet{rcm+03}, all of
the flux densities were measured or estimated to be $\gapp0.1$ Jy at
12-mm. $N_{\rm HI}$ is the neutral atomic hydrogen column density
(\scm) and $z_{\rm abs}$ is the redshift of the DLA,
respectively. $\nu_{\rm obs}$ is the observed frequency of the HCO$^+$
$0\rightarrow1$ line and $S$ is the flux density of the background
quasar previously measured at 12-mm.}
\begin{tabular}{@{}l c l c c @{}} 
\hline 
QSO & $\log_{10} N_{\rm HI}$ & $z_{\rm abs}$& $\nu_{\rm obs}$& $S$ \\ 
\hline
0201+113 &21.3  &  3.38639$^7$ & 20.333 & 0.6$^{a,d,e}$ \\
0201+365 & 20.4 &  2.4614$^4$ & 25.770 & 0.2$^*$ \\
0335--122 & 20.8 & 3.178$^9$ &  21.35 & 0.14$^e$ \\
0336--017 & 21.2 &  3.0619$^4$ & 21.958 & 0.15$^d$/0.18$^e$ \\
0405--331 & 20.6 & 2.570$^9$ & 24.98 & 0.72$^a$ \\
0528--250 & 21.2 & 2.8110$^5$ & 23.403 & 0.35$^e$ \\
0537--286 &20.3 & 2.974$^9$ & 22.43 & 1.66$^a$/0.6$^d$/0.9$^e$ \\
0913+003 & 20.3 & 2.774$^9$ &23.82 & 0.1$^e$ \\
1017+109 & 19.9 &  2.380$^2$ & 26.39 &  0.2$^*$ \\
1021--006 & 19.6 &  2.398$^2$ & 26.25 & 0.15$^c$/0.2$^e$ \\
1251--407 & 20.6 &  3.533$^9$ &19.68 & 0.06$^e$\\
...&20.3 &  3.752$^8$ &18.77 & ... \\
1354--107 &  20.4 & 2.501$^9$ & 25.48 &0.05$^e$ \\
...& 20.8 & 2.996$^8$ & 22.49 & ...\\
1402+044 & 20.2 &  2.485$^1$ &  25.59 & 0.6$^b$ \\
...&20.2 &  2.688$^1$ &23.32& ... \\
...&19.9 & 2.713$^1$ & 24.02& ... \\
1418--064 & 20.4 &  3.449$^9$ & 20.05& 0.3$^e$ \\
1614+051 & 20.4 & 2.52$^6$ & 25.3 & GPS \\
2131--045 & 20.0 &  3.27$^8$ & 20.9 & $<0.06$$^e$ \\
2342+342 & 21.2 &    2.908$^3$ &  22.82 &0.1$^*$\\
\hline
\end{tabular}
\label{gbt}
{References: (1) \citet{twl+89}, (2) \citet{lwt+91}, (3) \citet{wkb93}, (4) \citet{wlfc95}, (5) \citet{lsb+96}, (6) \citet{sw00}, (7) \citet{epss01}, (8) \citet{psm+01}, (9) \citet{eyh+02}.\\
($a$) ATCA calibrator, ($b$) \citet{knk+99}, ($c$) \citet{ttm+98}, ($d$) \citet{cmwp02}, ($e$) \citet{rcm+03}. Note that some flux densities$^*$ 
were estimated from neighbouring frequencies (see \protect\citealt{cwbc01})
and that 1614+051 is a Gigahertz Peaked Spectrum source.}
\end{minipage}
\end{table}

Apart from the sources listed in Table \ref{gbt}, there are only 3
other such DLAs known which have millimetre transitions redshifted to
$\approx12$-mm: Q 0438--436, RX J1028.6-0844 and LBQS
1213+0922. Although strong ($S\approx3$ Jy at 12-mm), the first source
is far too south to observe with the GBT, and the latter two were
estimated to have flux densities too low to yield good optical depth
limits within a reasonable integration time. Therefore, using the GBT we hoped for a
near complete survey of 12-mm absorption in known DLAs, although due
to time constraints we did not observe some of the sub-DLAs \citep{twl+89,lwt+91}
towards 1017+109, 1021--006 and 1402+044, nor the $N_{\rm
HI}=4\times10^{20}$ cm$^{-2}$ absorber at $z_{\rm abs}=3.533$ towards
1251--407 \citep{eyh+02}.

The observations were performed in April 2003, during several
consecutive days. Typically, under acceptable observing conditions,
the system temperature was 40--100 K. At the time of observing the
wide bandwidth spectrometer was unfortunately unavailable and so we
used the digital spectral processor with an observing bandwidth of
40~MHz, over 1024 channels.  This gave a redshift coverage of $\Delta
z_{\rm abs}\approx\pm0.004$, sufficiently wide to cover the absorption
redshift uncertainty in most of the DLAs. However, the optical
redshifts of 1614+051 and 2131--045 are poorly constrained (Table
\ref{gbt}) and so we observed these with five and three overlapping
bands, respectively. Dual polarization was used to optimise the
sensitivity and beam switching facilitated background emission
removal.

The data were reduced using the AIPS++ single dish package and
baselines were removed from the spectra by fitting a polynomial
function. The continuum levels were measured either during GBT
pointing scans of the brightest sources or with the Very Large Array
(VLA) in its most compact (D) configuration. These latter data were
taken in May 2003, and thus the continuum levels might have altered
due to quasar variability, although, as seen from Table \ref{sum},
these appear to be stable over the 9--11 month period since the ATCA
observations. For sources in which no continuum levels were
successfully measured within a month of the GBT observations, flux
densities were taken or estimated from the literature (see Table
\ref{sum}).

Also in May 2003 HCN $0\rightarrow1$ at $z_{\rm abs}=2.713$
towards 1402+044 was observed with the H22 receiver on the Nobeyama
45-m during less than ideal weather conditions during observations
for a different (3-mm) project \citep{cwmk03}. The 250 MHz AOS-W spectrometer (0.25
MHz channels) gave a redshift coverage of $\Delta z_{\rm abs}\approx\pm0.02$.

\subsection{3-mm observations}

In addition to the GBT survey, follow-up CO observations of candidate
molecular absorption detections \citep{cwn+02} were carried out using
the BIMA array (Table \ref{bima}). The spectrometer was configured in mode
5, with a central high resolution 100 MHz wide window (0.78 MHz
channels) flanked by two lower resolution 200 MHz wide windows (3.125
MHz channels), giving $\Delta z_{\rm abs} \approx\pm0.003 - 0.007$ for $z\sim0.2
- 2$. Some overlap was added in order to eliminate edge channels and
ensure the proper matching of the windows. The antennas where tuned to
the redshifted frequency of the chosen CO transition
\citep{lov92}, using a redshift that was the average of that of the
reported candidate absorption and the corresponding H{\small I}
absorption ($\Delta z_{\rm abs}\leq0.001$ in all cases). To correct for possible
spurious absorption detections due to RF or IF passband features, we
interleaved observations of a passband calibrator (a strong unresolved
radio source) with those of the target quasars every hour.  The total
integration time on the calibrator is such that the noise introduced
by the passband calibration is negligible. Since the data could be
self-calibrated, no observations of phase calibrators were
performed. The data were reduced using the MIRIAD interferometry
reduction package. The visibility cubes were imaged using natural
weighting corrected by system temperature, the resulting images were
cleaned, and spectra of the central positions were obtained. The final
passband-corrected spectra were produced by dividing the spectrum of
the source by that of its calibrator, then restoring the continuum
level by multiplying by the mean value of the calibrator. The
resulting spectra, along with the GBT results, can be viewed at
{\bf synergy url here please}
\begin{table}
\centering
\begin{minipage}{68mm}
\caption{The BIMA array targets. The parameters are
as given in Table \ref{gbt} where $\nu_{\rm obs}$ is the observed frequency
of the CO $0\rightarrow1$ line ($2\rightarrow3$ for 0458--020) and $S$
is the flux density of the background quasar previously measured at 3-mm.}
\begin{tabular}{@{}l c l c c @{}} 
\hline 
QSO & $\log N_{\rm HI}$ & $z_{\rm abs}$& $\nu_{\rm obs}$& $S$ \\ 
\hline
0248+430 & 21.6 & 0.3939$^2$ & 82.66  & $0.32$$^a$\\
0458--020 & 21.7 & 2.0399$^1$ & 113.76& $1.4-2.6$$^b$  \\
0738+313 & 21.3 & 0.2212$^2$ & 94.39 &  $0.27$$^a$  \\
\hline
\end{tabular}
\label{bima}
{References: (1) \citet{wlfc95}, (2) \citet{rt00}. ($a$) \citet{sjss95},
($b$) \citet{tlv00}.}
\end{minipage}
\end{table}


\section{Results and Discussion}

\subsection{Search results}

In Table \ref{sum} we show the best previously published optical
depths together with our new results, for which, despite the improved
limits, there are no absorption features of $\geq3\sigma$ over $\geq3$ \kms,
the full resolution of the BIMA array observations. For all of
the limits we estimate the $3\sigma$ upper limits on the total column
density of each molecule from
\begin{equation}
 N_{\rm mm}=\frac{8\pi}{c^3}\frac{\nu^{3}}{g_{J+1}A_{J+1\rightarrow J}}\frac{Qe^{E_J/kT_x}}{1-e^{-h\nu/kT_x}}
\left.\int\right.\tau dv,
\end{equation}
where $\nu$ is the rest frequency of the $J\rightarrow J+1$
transition, $g_{J+1}$ and $A_{J+1\rightarrow J}$ are the statistical
weight and the Einstein A-coefficient\footnote{These are taken from
\citet{cklh95,cms96} or derived from the dipole moment
(e.g. \citealt{rw00}).} of the transition, respectively, $Q =
\sum^{\infty}_{J=0}g_{J}~e^{-E_J/kT_x}$ is the partition
function\footnote{The energy of each level, $E_J$, is obtained from
the JPL Spectral Line Catalog \citep{ppc+98}.}, for the excitation
temperature, $T_x$, and $\int\tau dv$ is the $3\sigma$ upper limit
of the velocity integrated
optical depth of the line\footnote{$\int\tau dv\approx1.06\,\tau_{\rm
mm}\times{\rm FWHM}$ for a Gaussian profile, where $\tau_{\rm mm}$ is
the $3\sigma$ peak optical depth limit (Table \ref{sum}).}.

Since the derived optical depth limits depend upon the r.m.s. noise and
thus the velocity resolution (see \citealt{cwn+02}), in our previous
articles we presented all of the optical depth limits normalised to a
spectral resolution of 1 \kms, the finest resolution typical of most
current wide-band spectrometers.  Previously, we also quoted column
density upper limits per unit \kms~ line-width. However, in order to
use our limits to constrain molecular fractions at low redshift
(Section 3.2), we shall now adopt a line-width (FWHM) of 10 \kms,
which is close to those of the 4 known systems
(e.g. \citealt{wc96a}). This is done for a resolution of the same
value and so the column density limits, which are calculated for an
excitation temperature\footnote{Due to the increase of CMB temperature
with redshift, $T_{\rm CMB}=2.73 (1+z_{\rm abs})$, $T_{x}$ increases
to 20 K at $z_{\rm abs}=3.75$, the highest redshift of our sample. The
main effect of this is to increase the HCO$^+$ $0\rightarrow1$ column
density estimates by a factor of $\leq4$ in comparison to a constant
value of 10 K.} of 10 K at $z_{\rm abs}=0$, represent a one channel
$3\sigma$ detection of a $10$ \kms ~wide line.
\begin{table*}
 \centering
 \begin{minipage}{162mm}
\caption{Summary of published searches for molecular tracer absorption
in DLAs and sub-DLAs. $\nu_{\rm obs}$ is the approximate observed frequency (GHz),
$V$ is the visual magnitude of the background quasar and $S$ is the
flux density (Jy) at $\nu_{\rm obs}$ which, unless flagged, is the
value obtained during the actual observations (the limits are
$1\sigma$ and blanks in this field indicate that no value is given in
the literature). $N_{\rm HI}$ (\scm) is the DLA column density from
the Lyman-alpha line and $\tau_{\rm 21~cm}$ is the normalised peak
optical depth of the redshifted 21-cm line (see
\protect\citealt{cmp+03} for details). The optical depth of the
relevant millimetre line is calculated from $\tau=-\ln(1-3\sigma_{{\rm
rms}}/S)$, where $\sigma_{{\rm rms}}$ is the r.m.s. noise level. Since
$\sigma_{{\rm rms}}$ is dependent on the spectral resolution, we take
the various published values and recalculate $\sigma_{{\rm rms}}$ at a
resolution of 10 \kms ~($\tau_{\rm mm}$) and quote only the best
existing limit. For all optical depths, $3\sigma$ upper limits are
quoted and ``--'' designates where $3\sigma>S_{{\rm cont}}$, thus not
giving a meaningful value for this limit. The penultimate column gives
the corresponding column density (\scm) per 10 \kms ~channel (see main
text).}
\begin{tabular}{@{}l c c r c c c c c c c @{}}
\hline
DLA & $z_{\rm abs}$ & Transition & $\nu_{\rm obs}$ & $V$ & $S$ & $N_{\rm HI}$ & $\tau_{\rm 21~cm}$ &$\tau_{\rm mm}$ & $N_{\rm mm}$ & Ref.\\
\hline
0201+113 & 3.38639 & HCO$^+$ $0\rightarrow1$ & 20.3 & 19.5 &0.47 & $2\times10^{21}$ &  $\leq0.09$ &$<0.007$ & $<1\times10^{12}$ & 10\\
0201+365 & 2.4614 & HCO$^+$ $0\rightarrow1$ & 25.8 & 17.9 &0.09$^V$  & $3\times10^{20}$  & & $<0.1$ & $<1\times10^{13}$ & 10\\ 
0235+164 & 0.52400 & CO $0\rightarrow1$& 75.6 &15.5 & 2.5& $4\times10^{21}$ & $0.64$ &$<0.02$&$<1\times10^{15}$ & 2\\
... &0.52398 & CO $1\rightarrow2$ & 151.3& ... & 1.27& ... & ... & $<0.03$& $<1\times10^{15}$& 6\\
...	&  ...& HCO$^+$ $3\rightarrow4$ & 234.1 & ... & 0.75&... &...  & $<0.1$  &$<7\times10^{12}$ & 6\\
...& 0.523869& CS $2\rightarrow3$ & 96.4 &... & $1.7$ &...&  ...&$<0.1$  &$<4\times10^{13}$ & 8\\
0248+430 & 0.3939& CS $2\rightarrow3$ & 105.4 &17.7 &$<0.2$ &$4\times10^{21}$ &  0.2& $<0.4$&$<1\times10^{14}$ & 8\\
...& 0.394 &  CO $0\rightarrow1$& 82.7 & ... & 0.21    &...&  ...&  $<0.4$ &$<3\times10^{16}$  & 10\\
0335--122 & 3.178 &  HCO$^+$ $0\rightarrow1$ & 21.4 & 20.2 & 0.13$^{A,V}$& $6\times10^{20}$  &$<0.008$ &$<0.03$&  $<6\times10^{12}$& 10\\ 
0336--017 & 3.0619 & HCO$^+$ $0\rightarrow1$ &22.0 &18.8 & 0.15$^A$& $2\times10^{21}$ &$<0.007$ &$<0.03$& $<5\times10^{12}$& 10\\
0405--331 &2.570 & HCO$^+$ $0\rightarrow1$ & 25.0 & 19.0 & 0.53&$4\times10^{20}$ & &$<0.02$&$<3\times10^{12}$ & 10\\
0458--020 & 2.0397  & HCO$^+$  $2\rightarrow3$  & 88.0 &18.4 & & $5\times10^{21}$ &0.3  & $<0.3$& $<1\times10^{13}$& 7\\
...	&   2.0399 &... & 88.0 & ...&$1.3$& ... &... & $<0.07$ &$<3\times10^{12}$ & 8\\
...	& 2.03937 &CO $0\rightarrow1$ & 37.9 &  ...& $\approx0.8$ & ...&  ... & $<0.01$ &$<1\times10^{15}$ & 4\\
...	&  2.0398 & CO $2\rightarrow3$ & 113.8 &... &0.53 & ...&  ... &  $<0.1$& $<6\times10^{15}$ & 10\\
...	&  2.0397 & CO $2\rightarrow3$ & 113.8 &... &0.53$^a$ & ...&  ... &$<0.6$&  $<4\times10^{16}$& 5\\
...	&   2.0399 & CO $3\rightarrow4$& 151.7 &... & $0.4$& ...& ... &$<0.4$ & $<4\times10^{16}$ & 8\\
...& 2.04 & H$_2$CO $1_{10}\rightarrow1_{11}$ & 1.6 & ...& & ...&...&$<0.01$ &  & 3\\
0528--2505 & 2.1408 & CO $2\rightarrow3$ &110.1 &19.0 & $\approx0.2^b$ &$4\times10^{20}$ & $<0.3$ & $<1.0$& $<6\times10^{16}$& 5\\
... & ... & HCO$^+$ $0\rightarrow1$ &23.4 &  ...& 0.3$^{A,V}$ & ... & ... & $<0.02$&$<3\times10^{12}$ & 10\\
0537--286 & 2.974 & HCO$^+$ $0\rightarrow1$ & 22.4 &19.0 & 0.58& $2\times10^{20}$ &$<0.007$ &$<0.02$ &$<3\times10^{12}$ & 9\\
0738+313 &  0.2212 & CO $0\rightarrow1$ & 94.4 &16.1 &$0.48$ & $2\times10^{21}$ & $\approx0.07$& $<0.04$& $<3\times10^{15}$  & 10\\
0827+243 & 0.5247 & CS $2\rightarrow3$ & 96.4 &17.3  &$2.7$ & $2\times10^{20}$ &0.007  &$<0.09$ & $<3\times10^{13}$& 8\\
08279+5255  & 2.97364& HCO$^+$ $0\rightarrow1$ & 44.9 &15.2 & &$1\times10^{20}$ &  & \multicolumn{2}{c}{{\it No 7 mm flux available}}   & 8\\
...	&  ... & CO $2\rightarrow3$ & 87.0 &... & $<0.1$ &... & ... & & & 8\\
0834--201 & 1.715 & HCO$^+$  $2\rightarrow3$ &98.6 &  18.5& $1.7$ & $3\times10^{20}$ & & $<0.1$&$<4\times10^{12}$  & 8\\
...	&  ... &HCO$^+$  $3\rightarrow4$& 131.4 &... & & ...&  ...& $<0.2$&$<1\times10^{13}$ & 7\\
...	&  ... &  CO $3\rightarrow4$& 169.8 &... & $0.9$& ...&  ...&$<0.7$ &$<8\times10^{16}$  & 8\\
0913+003 & 2.774 & HCO$^+$ $0\rightarrow1$ & 23.8 & -- & $\approx0.17$$^{A,V}$ & $2\times10^{20}$ &  &$<0.04$ & $<6\times10^{12}$ &10\\ 
1017+1055 & 2.380& CS $2\rightarrow3$ & 43.5 &17.2 & & $8\times10^{19}$ &  & \multicolumn{2}{c}{{\it No 7 mm flux available}}  & 8\\
...	&  ... & CO $2\rightarrow3$ & 102.3 &... &$<0.2$ &  ...&  ... & --& --&   8\\
1215+333 & 1.9984 & CO $2\rightarrow3$ & 115.3 &18.1 & & $1\times10^{21}$ &   &   \multicolumn{2}{c}{{\it No 3 mm flux available}}  & 5\\
1229--021 & 0.3950 & CO $0\rightarrow1$  & 82.6 &16.8  &$0.2$ & $1\times10^{21}$ &0.05  & $<0.7$& $<5\times10^{16}$& 8\\
...	&  ... & CO $1\rightarrow2$& 165.3 &... &$<0.1$ &  ...&  ...& --& --& 8\\
...	& 0.39498  & CO $1\rightarrow2$& 165.3 &... &$0.11$ &...&  ...&-- &-- &6\\
1251--407 & 3.752 &   HCO$^+$ $0\rightarrow1$ & 18.8 &23.7 & $\approx0.1$$^{A,V}$ & $2\times10^{20}$ & &$<0.09$& $<2\times10^{13}$& 10\\
1328+307 & 0.69215 & HCO$^+$  $1\rightarrow2$ & 105.4 &17.3  &0.50 & $2\times10^{21}$ &0.02  & $<0.2$& $<6\times10^{12}$& 6\\
...	&  ... & CS $2\rightarrow3$ & 86.9 &... & $1.0$ &... & .... & $<0.3$& $<8\times10^{13}$ &  8\\
...	&  ... & CO $1\rightarrow2$ & 136.2 &... & 0.39& ...& ... &$<0.3$ &$<1\times10^{16}$ & 6\\
...	&  ... & CO $2\rightarrow3$ & 204.4 &... & 0.27& ...& ... &$<0.5$ & $<3\times10^{16}$& 6\\
1331+170 & 1.7764 & CO $0\rightarrow1$ & 41.5 &16.7 & 0.5& $3\times10^{21}$ & 0.02 &$<0.4$ & $<5\times10^{16}$ & 1\\
...& 1.7755 & ...&41.5 &  ... & ...& ...& ... &$<0.4$ &$<5\times10^{16}$ & 1\\
1354--107 & 2.501 & HCO$^+$ $0\rightarrow1$ & 25.5 & 19.2 &$\approx0.07$$^{A,V}$  &  $3\times10^{20}$ & $<0.015$ & $<0.08$&$<1\times10^{13}$ & 10\\
... &  2.996 & ...& 22.5 & ... &...  & ... & ... & $<0.1$& $<2\times10^{13}$& 10\\
1402+044 &  2.713 & HCO$^+$ $0\rightarrow1$ & 24.0 & 19.8 & 0.21$^*$ & $8\times10^{19}$ &  &$<0.008$& $<1\times10^{12}$ & 10\\
... &... & HCN  $0\rightarrow1$ & 23.9 & ... & 0.6 & ... & ...& $<0.02$& $<8\times10^{12}$ & 10\\
1418--064 & 3.449 &  HCO$^+$ $0\rightarrow1$ & 20.1 & 18.5 & 0.14 & $3\times10^{20}$ &   &$<0.02$ & $<4\times10^{12}$ & 10\\
1451--375 &0.2761 & HCO$^+$ $1\rightarrow2$ & 139.8 & 16.7 &$0.6$ & $1\times10^{20}$ & $<0.007$ & $<0.2$ &$<7\times10^{12}$  & 8\\
...	&  ... &CO $0\rightarrow1$  &90.3 & ... & $1.2$ & ...& ... & $<0.1$&$<8\times10^{15}$& 8\\
\hline
\end{tabular}
\label{sum}
{References: (1) \citet{tsi+84}, (2) \citet{tnb+87}, (3)
  \citet{bwl+89}, (4) \citet{tn91}, (5) \citet{wc94}, (6) \citet{wc95}, (7)
  \citet{wc96b}, (8) \citet{cwn+02}, (9) \citet{cmwp02}, (10) This
  paper.\\ Flux densities: $^A$ATCA June \& August 2002
  \citep{cmwp02,rcm+03}, $^V$VLA May 2003. Where the flux densities
  could not be obtained from the data/article -- $^a$the BIMA array value,
  $^b$interpolated between 11 GHz and K-band, $^c$interpolated between
  radio and 0.5--10 keV, $^d$extrapolated from 0.4 and 5 GHz. $^*$This
  observation of 1402+044 is from preliminary (February 2003)
  GBT observations using the wide band spectrometer.}
\end{minipage}
\end{table*}
\begin{table*}
\addtocounter{table}{-1}
 \centering
 \begin{minipage}{167mm}
\caption{{\it Continued}}
\begin{tabular}{@{}l c c r c c c c c c c @{}}
\hline
DLA & $z_{\rm abs}$ & Transition & $\nu_{\rm obs}$ & $V$ & $S$ & $N_{\rm HI}$ & $\tau_{\rm 21~cm}$ &$\tau_{\rm mm}$ & $N_{\rm mm}$ & Ref.\\
\hline
1614+051 &  2.52 & HCO$^+$ $0\rightarrow1$ & 25.3 & 19.5 & $\sim0.3$$^c$ & $3\times10^{20}$ & & $<0.03$&$<4\times10^{12}$ & 10\\
2131--045 & 3.27 & HCO$^+$ $0\rightarrow1$ & 20.9 & 20.0 &  $<0.06^A$ &$1\times10^{20}$ & & --&--  & 10\\
2128--123 & 0.4298 & CS $2\rightarrow3$ & 102.8 & 15.5 & 0.75 & $2\times10^{19}$& $<0.003$ & $<0.04$ &$<1\times10^{13}$ & 7\\ 
...	&  ... & CS $3\rightarrow4$ & 137.1 & ... & 0.70 &  ...&   ...& $<0.05$ &$<2\times10^{13}$ & 7\\ 
...	&  ... & CO $1\rightarrow2$ & 161.2 & ... & 0.5 & ...&   ...& $<0.2$ &$<6\times10^{15}$ & 7\\ 
...	&  ... & CO $2\rightarrow3$ & 241.9 & ... & 0.4 & ...&   ...& $<0.2$ &$<1\times10^{16}$ & 7\\ 
2136+141 & 2.1346 & HCO$^+$ $2\rightarrow3$&85.4 & 18.9  & 0.59& $6\times10^{19}$&   &$<0.1$ &$<4\times10^{12}$ & 6\\
...	&  ... &  CO $2\rightarrow3$ & 110.3 &... &0.27 & ...&   ...&$<0.2$  &$<9\times10^{15}$ &7\\
...	&  ... &  CO $3\rightarrow4$ &147.1 & ... & 0.21& ...&   ...&$<0.2$  & $<2\times10^{16}$& 7\\
...	&  ...&  CO $5\rightarrow6$ & 220.6 &... & 0.16 &...&  ... &$<0.6$ & $<8\times10^{17}$ &  7\\
2342+342 & 2.908 &  HCO$^+$ $0\rightarrow1$ & 22.8 & 18.4 & $\sim0.1$$^d$ &$2\times10^{21}$  &$<0.03$ & $<0.04$& $<7\times10^{12}$ & 10\\
\hline
\end{tabular}
\end{minipage}
\end{table*}

\subsection{Deriving molecular fraction limits}

Despite our improved limits there has yet to be a detection of a
rotational molecular transition in a DLA. In many cases the limits far
exceed those required to detect the 4 known redshifted millimetre
absorbers (see below, Figs.  \ref{co} and \ref{hco}). These systems
have molecular fractions of $f\equiv\frac{2N_{{\rm H}_{2}}}{2N_{{\rm
H}_{2}}+N_{{\rm HI}}}\approx0.3-1.0$ (e.g. \citealt{cw98b}), c.f.
$f\sim10^{-7}-10^{-2}$ for the 10 DLAs in which molecular hydrogen has
been detected at UV and optical wavelengths (\citealt{rbql03} and
references therein). In this section we investigate the possibility of
detecting molecular tracers in systems with such low molecular
fraction.

\begin{figure}
\vspace{5.3cm}
\includegraphics{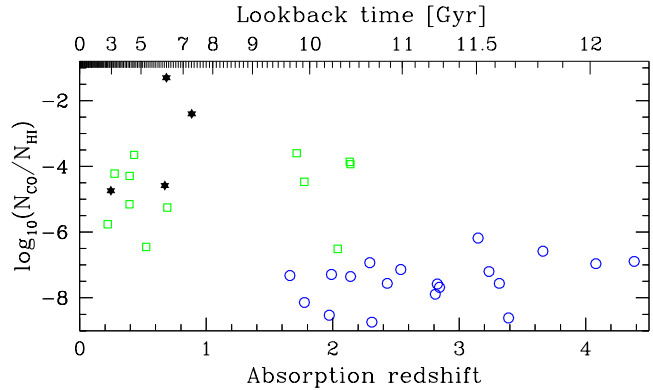}
\caption{The normalised CO column density versus the absorption
redshift. The unfilled markers designate upper limits -- squares for
the millimetre searches and circles for the optical searches. The
filled stars represent the 4 known millimetre absorption systems. The
lookback time is for $\Omega_{\rm m} = 0.30, \Omega_{\Lambda}=0.70,
H_0 = 70\,{\rm km\,s}^{-1}\,{\rm Mpc}^{-1}$.}
\label{co}
\end{figure}
In Fig. \ref{co} we show the CO rotational (Table \ref{sum}) and
electronic \citep{gbwb97,lsb99,psl02} column density limits along with
the values for the 4 known systems. While this illustrates the
sensitivity to detecting absorption in the known systems, it clearly
shows that the radio searches are considerably less sensitive than the
optical CO searches\footnote{Although the deep integrations of
0235+164 and 0458--020 with the NRO 45-m telescope \citep{tnb+87,tn91}
do approach the optical limits.}. They do, however, complement the
optical results in giving limits at redshifts of $z_{\rm abs}\lapp1.8$
(below the atmospheric cut-off for direct optical \MOLH ~detections).
In Fig. \ref{hco} we show the HCO$^+$ column density limits.
\begin{figure}
\vspace{5.3cm}
\includegraphics{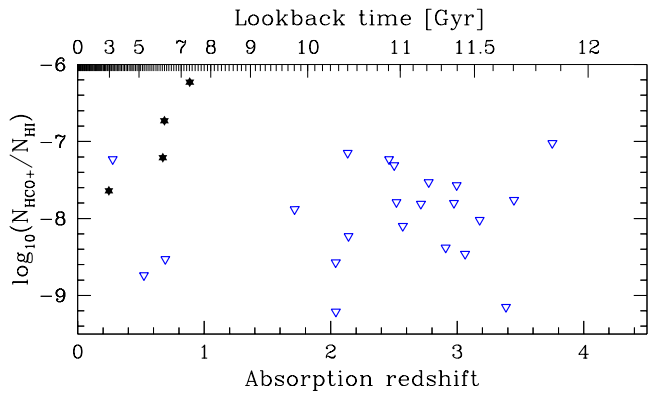}
\caption{The normalised HCO$^+$ column density versus the absorption
redshift. Again the stars represent the 4 known systems and the inverted triangles are the
DLA upper limits (Table \ref{sum}).}
\label{hco}
\end{figure}
Again we see that our limits are more than sufficient to detect
HCO$^+$ in the 4 known systems \citep{wc95,wc96,wc96b,wc97}, with
two cases having been searched to at least an order of magnitude
better than that required to detect the known systems at low redshift.

Although we cannot match the sensitivity to molecular absorption
   provided by the optical results, we can nevertheless attempt to
   derive constraints on the molecular fraction for absorption systems
   at low redshift. The major obstacle in this is determining which
   $N_{{\rm CO}}$--$N_{{\rm H}_{2}}$ conversion ratio to apply: Unlike
   HCO$^+$ (discussed later), the ratio may vary from $N_{{\rm
   H}_{2}}\sim10^6 N_{{\rm CO}}$ for diffuse gas \citep{ll00a} to
   $N_{{\rm H}_{2}}\sim10^4 N_{{\rm CO}}$ for dense, dark Galactic
   clouds. This latter value is applied to the 4 known millimetre
   absorbers to yield molecular fractions from rotational lines (see
   \citealt{wc98})\footnote{Since all of these sources absorb at low
   redshift, $z_{\rm abs}\leq0.87$, and have $V\gapp20$ there is
   currently little chance of detecting the \MOLH ~line
   directly.}. However, DLAs have lower metallicities than Galactic
   systems and are less visually obscured than the 4 known systems,
   thus casting doubt on whether applying this conversion ratio is
   justified. 

\begin{figure}
\vspace{9.0cm}
\includegraphics{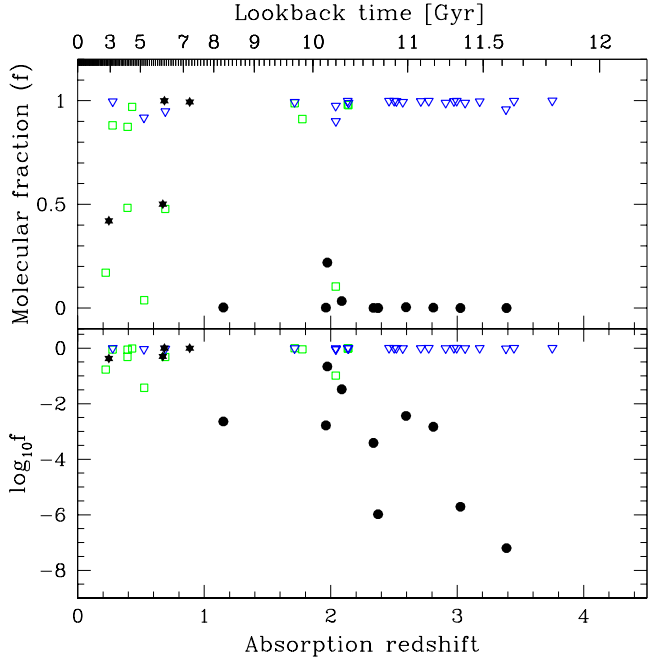}
\caption{Molecular fractions measured in redshifted absorbers. Again
the unfilled symbols represent the upper limits from the millimetre
searches, $N_{{\rm H}_{2}}=10^{4.75} N_{{\rm CO}}$ and $T_x = 10$ K
at $z_{\rm abs}=0$, with the squares representing CO and the inverted
triangles HCO$^+$. The circles represent the H$_2$-bearing DLAs and
the stars the 4 known millimetre absorption systems ($N_{{\rm
H}_{2}}=10^{4} N_{{\rm CO}}$ and  $T_x = 10$ K, $\forall z_{\rm abs}$). A log version of the plot
is also shown by means of a detail of the H$_2$-bearing systems.  This
clearly illustrates the evolution of $f$ described by \citet{cwmc03}.}
\label{frac}
\end{figure}

If we extrapolate the metallicity
   ([M/H]\footnote{Defined as the heavy element abundance with respect
   to hydrogen, relative to that of the solar neighbourhood: ${\rm
   [M/H]} \equiv \log_{10}[N({\rm M})/N({\rm H})] - \log_{10}[N({\rm
   M})/N({\rm H})]_\odot$.}) evolution of the general DLA population\footnote{Due to many of the sources in Table \ref{sum} having low redshifts
and/or high visual magnitudes, there are only metallicity measurements
for half of these sources, all of which belong to the sample of
\citet{pgw+03}. Therefore, by definition these follow the same
metallicity evolution of the general DLA population rather than that
of the \MOLH-bearing DLAs \citep{cwmc03}.}
according to $[{\rm
M/H}]\approx-0.26 z_{\rm abs}$ \citep{kf02,pgw+03},
\begin{figure}
\vspace{5.3cm}
\includegraphics{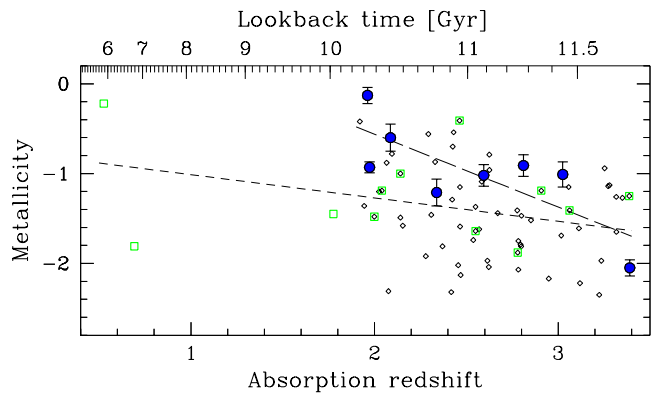}
\caption{The metallicity evolution of DLAs. The filled circles
represent the DLAs which exhibit H$_2$ absorption with the
least-squares fit shown \protect\citep{cwmc03}, the small unfilled
diamonds represent the metallicity measurements of 60 DLAs between
$1.9 < z_{\rm abs} < 3.4$ (i.e. the general population), again with
the fit shown \citep{pgw+03}, and the unfilled squares are the DLAs
searched for millimetre absorption (Table \ref{sum}). }
\label{metal}
\end{figure}
we obtain $[{\rm M/H}]\approx-0.75$ at $z_{\rm abs}\sim0$
(Fig. \ref{metal}). At zero redshift this is still significantly lower
than solar values suggesting that, in addition to lower enrichment at
high redshift, there may be appreciable dust depletion of the metals,
as in dense Galactic clouds (e.g. \citealt{ss96}), although this may
be more relevant to the \MOLH-bearing DLAs where $f\gapp10^{-2}$
at $z_{\rm abs}\lapp1.8$ \citep{cwmc03}. Moreover,
\citet{lps03} suggest that the large depletion factors in DLAs
indicate a significant dust content.  Therefore, a fair compromise
between the diffuse/dark cases may be to assume the conversion ratio
for dark clouds but scaled by the abundance of metals from which the
tracer molecules form, i.e. $N_{{\rm H}_{2}}\approx10^{4.75} N_{{\rm
CO}}$ at $z_{\rm abs}\sim0$. Since the metallicity decreases with
increasing redshift, we may expect the conversion of tracer to
molecular hydrogen column density to scale accordingly. We therefore
apply the ratio of $N_{{\rm H}_{2}}\approx10^{0.26 z_{\rm abs} + 4.75}
N_{{\rm CO}}$, based upon the metallicity evolution of the general DLA
population\footnote{Since the molecular hydrogen fraction shows a
strong anti-correlation with redshift for the \MOLH-bearing DLAs
(\citealt{cwmc03} and Fig. \ref{frac}), we may expect less depletion
and thus relatively higher metallicities. Fig. \ref{metal} does appear
to suggest, however, that the metallicity is dominated by poorer
chemical enrichment at higher redshifts.}. This gives $N_{{\rm
H}_{2}}\sim10^5 N_{{\rm CO}}$ at $z\sim1$ rising to $N_{{\rm
H}_{2}}\sim10^6 N_{{\rm CO}}$ at $z\sim5$.  From optical CO limits,
\citet{bcf87} and \citet{cbf88} have previously noted that the carbon
to hydrogen column density ratio is a tenth of the Galactic value at
$z_{\rm abs}=1.7$ and 2.3. 

In addition to CO, we can use the HCO$^+$ limits, which could be a
better choice of tracer, since a near constant ratio of $N_{{\rm
HCO^+}}=2-3\times10^{-9} N_{{\rm H}_{2}}$ is found over various
regimes in the Galaxy \citep{ll00}. Applying the conversion thus
suggests that $N_{{\rm H}_{2}}\sim10^9 \rightarrow 10^{10} N_{{\rm
HCO^+}}$ at $z\sim1\rightarrow 5$, i.e. $N_{{\rm
H}_{2}}\approx10^{0.26 z_{\rm abs} + 9.34} N_{{\rm HCO}^+}$. In
Fig. \ref{frac} we show the derived molecular fractions
of the CO and HCO$^+$ limits together with those of the H$_2$-bearing
DLAs\footnote{Versions of Fig. \ref{frac} for $N_{{\rm H}_{2}}=10^6
N_{{\rm CO}}$ and $10^4 N_{{\rm CO}}$ can be viewed at
\bf synergy url here please}.  From this we
see that the high conversion ratio and excitation above the
$0\rightarrow1$ transition make the HCO$^+$ molecule insensitive with
current telescopes to $f\lapp1$: Although in many cases our GBT
observations yield significantly better limits than previously (Table
\ref{sum}), these are all at high redshift and therefore convert less
favourably to molecular fractions due to the steep metallicity
evolution. Thus, we obtain near constant limits to the fraction across
the entire redshift range searched for HCO$^+$.  Note that the optical
CO limits (Fig. \ref{co}) occupy the space between the HCO$^+$ limits
and the \MOLH-bearing DLAs ($1.7\lapp z_{\rm abs}\lapp4.4$, $-3 \lapp
\log_{10} f \lapp-0.5$) when the conversion is applied.

We do see however, based on our redshift dependent conversion factor,
that the millimetre CO searches generally give interesting (low
redshift) limits, with two of the limits within the range of the
H$_2$-bearing DLAs. This suggests that, were these DLAs
\MOLH-bearing\footnote{\MOLH ~is detected in $\lapp20\%$ of recent DLA
surveys \citep{lps03}, and so we perhaps expect a $\approx1/5$ chance
of having observed an \MOLH-bearing system.}, making the
$N_{{\rm CO}}$--$N_{{\rm H}_{2}}$ conversion reasonable, the CO
rotational transition could have been detected in these DLAs: 0235+164
($z_{\rm abs}=0.524$) and 0738+313 ($z_{\rm abs}=0.221$).
Unfortunately, there are no metallicity measurements available for
these, although \citep{pgw+03} use the value of
${\rm[M/H]}=-0.22\pm0.15$ for 0235+164 which comes from model
dependent X-ray measurements \citep{trp+03}. Several other models give
lower values than this, thus suggesting that this DLA may belong to
the general population (${\rm [M/H]}\approx-0.9$), although the value
quoted by \citet{pgw+03} could place it in either population
(Fig. \ref{metal}). The general population fit is, however, consistent
with the range of $-1.4 < {\rm [M/H]} < -0.4$ at $z_{\rm
abs}\approx0.4$ estimated by \citet{trp+03} and the fact that we did
not detect CO. Regarding the molecular fraction, the data are
sensitive to $\log_{10} f\sim-1.5$ for the general population
metallicity and $\log_{10} f\sim-2.1$ for the value used by
\citet{pgw+03}. This latter fraction shifts this DLA further into the
\MOLH-bearing DLA regime. 

For more diffuse clouds of low dust content our limits become weaker
with $f\geq0.4$, c.f. $\geq0.03$ previously, being obtained for
$N_{{\rm H}_{2}}\sim10^6 N_{{\rm CO}}$ and $T_x = 10$ K at $z_{\rm
abs}=0$, although in the diffuse gas case the column density limits
would improve slightly due to lower excitation temperatures (at best a
factor of $\approx0.5$ lower than given in Table \ref{sum})$^{13}$. That is,
the clouds being diffuse could also explain why no millimetre
absorption of CO was detected.

\subsection{Detecting millimetre lines in DLAs with the next generation of large radio telescopes}

Applying the conversion of tracer column densities described above, we
can ascertain the likelihood of detecting these molecular tracers in
DLAs with the next generation of large radio telescopes. For $N_{{\rm
H}_{2}}=10^{4.75} N_{{\rm CO}}$ and $T_x = 10$ K at $z_{\rm abs}=0$:
\begin{enumerate}
  \item The Atacama Large Millimeter Array
(ALMA)\footnote{http://www.eso.org/projects/alma/}: The 3-mm band of
this telescope will give CO $0\rightarrow1$ coverage to $z\leq0.34$.
At 10 \kms ~resolution, one hour of integration will give an r.m.s. of
$\approx0.3$ mJy. For a 0.3 Jy continuum flux (fairly typical in Table
\ref{sum}), this corresponds to a $3\sigma$ column density limit
of $N_{{\rm CO}}\sim10^{14}$ \scm ~per channel. Applying the conversion
at these redshifts gives $N_{{\rm H}_{2}}\sim10^{19}$ \scm ~thus giving
$f\sim10^{-3}-0.1$ for DLAs. If we extrapolate the molecular fractions
in  \MOLH-bearing systems back to $z\lapp0.3$ (Fig. \ref{frac}),
we see that such limits could well be sufficient to detect rotational CO in 
DLAs at low redshift.
        \item The Square Kilometre Array
	  (SKA)\footnote{http://www.skatelescope.org/}: With a tuning
	  range of 0.15 to 20 GHz this telescope can observe CO at
	  $z\geq4.8$. With the same parameters as above, an r.m.s. of
	  5 $\mu$Jy is reached giving $N_{{\rm CO}}\sim10^{12}$ \scm
	  ~per channel\footnote{Again, this is for a flux density of
	  0.3 Jy. Due to the antenna temperature dominating the system
	  at these low frequencies, an increased flux does not improve
	  the optical depth to the same extent as for the millimetre
	  observations. Therefore, the column density estimate is
	  fairly robust.}, thus rivalling current optical limits.  At
	  $z\sim5$, $N_{{\rm H}_{2}}\sim10^6 N_{{\rm CO}}$ giving
	  $f\sim10^{-4} - 10^{-2}$ for DLAs, i.e. the molecular
	  fraction range of 7 of the 10 \MOLH-bearing DLAs
	  (e.g. \citealt{rbql03}). Unfortunately at $z\gapp5$,
	  extrapolating the molecular fraction--redshift
	  anti-correlation, we expect $f\lapp10^{-12}$, thus making it
	  unlikely that redshifted millimetre transitions could be
	  detected\footnote{Note, however, that from Galactic studies,
	  \citet{lis02} suggests that at very low metallicities \MOLH
	  ~will still via form slow gas-phase processes, giving minimum
	  molecular fractions of $f\sim10^{-8}-10^{-7}$. This
	  suggests that current DLA molecular fraction measurements
	  may be close to the minimum possible value (see
	  \citealt{lps03,rbql03}).} with the SKA, despite its
	  superior sensitivity\footnote{Although \citet{cdk04} discuss
	  the possibility of detecting low redshift OH 18-cm lines in DLAs
	  with this telescope.}.
	  
	  \end{enumerate}

\section{Summary}

We have used the GBT and BIMA array to search for HCO$^+$ and CO
absorption in 19 damped Lyman-alpha absorption systems. Although in
many cases the sensitivity reached far surpasses that required to
detect the 4 known redshifted millimetre absorption systems, a
rotational millimetre-band transition has yet to be detected in a DLA. From optical
studies it is revealed that molecular fractions in DLAs are much lower
than those of the 4 known systems and, in those where \MOLH ~has been
detected, the \MOLH ~molecular fraction in DLAs decreases with increasing redshift.

This evolution is also apparent in DLA metallicities, which we use to
gauge the abundance of heavy elements available to form tracer
molecules. By scaling the \MOLH--CO \& --HCO$^+$ column density ratios
accordingly, we find:
\begin{itemize}
  \item Although the GBT results offer a significant improvement over
    previous HCO$^+$ searches, the high redshifts ($z\gapp2.5$) yield
    correspondingly large conversion ratios and excitation
    temperatures, i.e. $N_{{\rm H}_{2}}\gapp10^{10}N_{{\rm HCO^+}}$,
    rendering high redshift searches sensitive only to molecular
    fractions of $f\sim1$.  \item Since many of the CO searches are
    performed in the 3-mm band, these are generally at much lower
    redshift. So in addition to the $z\gapp1.7$ optical limits of
    electronic transitions, we can estimate lower redshift values from
    the rotational transitions. Although the statistics are small and
    the quality of the limits varied (i.e. some limits are only
    sensitive to $f\gapp0.9$), assuming $N_{{\rm
    H}_{2}}\approx10^{0.26 z_{\rm abs} + 4.75} N_{{\rm CO}}$, we
    obtain, for the general (non-\MOLH) DLA population, a $3\sigma$
    limit of $f\lapp0.03$ at $z=0.5$. For more diffuse clouds
    ($N_{{\rm H}_{2}}\approx10^{0.26 z_{\rm abs} + 6} N_{{\rm CO}}$)
    this increases to $f\lapp0.2$.

\end{itemize}
Finally, we discuss the possibility of detecting CO absorption with
the next generation of large radio telescopes in the context of
the metallicity evolution model. Although the SKA will reach
sensitivities of $\sim5~\mu$Jy after one hour of integration
(c.f. $\sim300~\mu$Jy with ALMA), ALMA is the most likely to detect
millimetre absorption in DLAs at low redshift due to its high
frequency capabilities.

\section*{Acknowledgments}
We wish to thank Carl Bignell and Nario Kuno for their invaluable
assistance in operating the Green Bank and Nobeyama 45-m telescopes,
respectively. Also, the anonymous referee, Harvey Liszt and Max
Pettini for their helpful comments, Chris Blake, Michael Burton, and
Matthew Whiting for their input, as well as the John Templeton
Foundation for supporting this work. SJC gratefully acknowledges
receipt of a UNSW NS Global Fellowship and MTM is grateful to PPARC
for support at the IoA under the observational rolling grant
(PPA/G/O/2000/00039). This research has made use of the NASA/IPAC
Extragalactic Database (NED) which is operated by the Jet Propulsion
Laboratory, California Institute of Technology, under contract with
the National Aeronautics and Space Administration.

\bsp

\label{lastpage}


\begin{thebibliography}{57}
\expandafter\ifx\csname natexlab\endcsname\relax\def\natexlab#1{#1}\fi

\bibitem[{{Black} {et~al.}(1987){Black}, {Chaffee} \& {Foltz}}]{bcf87}
{Black} J.~H., {Chaffee} F.~H., {Foltz} C.~B., 1987, ApJ, 317, 442

\bibitem[{{Briggs} {et~al.}(1989){Briggs}, {Wolfe}, {Liszt}, {Davis} \& 
  {Turner}}]{bwl+89}
{Briggs} F.~H., {Wolfe} A.~M., {Liszt} H.~S., {Davis} M.~M., {Turner}
  K.~L., 1989, ApJ, 341, 650

\bibitem[{{Chaffee} {et~al.}(1988){Chaffee}, {Black} \& {Foltz}}]{cbf88}
{Chaffee} F.~H., {Black} J.~H., {Foltz} C.~B., 1988, ApJ, 335, 584

\bibitem[{Chandra {et~al.}(1995)Chandra, Kegel, Roy,  Hertenstein}]{cklh95}
Chandra S., Kegel W.~H., Roy R. J.~L.,  Hertenstein T., 1995, A\&AS, 114,
  175

\bibitem[{Chandra {et~al.}(1996)Chandra, Maheshwari,  Sharma}]{cms96}
Chandra S., Maheshwari V.~U.,  Sharma A.~K., 1996, A\&AS, 117, 557

\bibitem[{{Combes} \& {Wiklind}(1998)}]{cw98b}
{Combes} F. {Wiklind} T., 1998, ESO Messenger, 91, 29

\bibitem[{Curran {et~al.}(2004{\natexlab{a}})Curran, Darling, 
  Kanekar}]{cdk04}
Curran S.~J., Darling J.~K.,  Kanekar N., 2004{\natexlab{a}}, in
 Carilli C. L. et al., eds, Astrophysics with the Square Kilometer 
 Array, Elsevier, in preparation

\bibitem[{Curran {et~al.}(2004{\natexlab{b}})Curran, Murphy, Pihlstr\"{o}m,
  Purcell, Webb,  Kanekar}]{cmp+03}
Curran S.~J., Murphy M. .~T., Pihlstr\"{o}m Y.~M., Purcell C.~R., Webb,
  J.~K.  Kanekar, N., 2004{\natexlab{b}}, MNRAS, in preparation

\bibitem[{Curran {et~al.}(2003)Curran, Murphy, Webb,  Pihlstr\"{o}m}]{cmwp02}
Curran S.~J., Murphy M.~T., Webb J.~K.,  Pihlstr\"{o}m Y.~M., 2003, MNRAS,
  340, 139

\bibitem[{Curran {et~al.}(2002{\natexlab{a}})Curran, Murphy, Webb,
  Rantakyr\"{o}, Johansson,  Nikoli\'{c}}]{cwn+02}
Curran S.~J., Murphy M.~T., Webb J.~K., Rantakyr\"{o} F., Johansson L.
  E.~B.,  Nikoli\'{c} S., 2002{\natexlab{a}}, A\&A, 394, 763

\bibitem[{Curran {et~al.}(2002{\natexlab{b}})Curran, Webb, Murphy, Bandiera,
  Corbelli,  Flambaum}]{cwbc01}
Curran S.~J., Webb J.~K., Murphy M.~T., Bandiera R., Corbelli E., 
  Flambaum V.~V., 2002{\natexlab{b}}, PASA, 19, 455

\bibitem[{Curran {et~al.}(2004{\natexlab{c}})Curran, Webb, Murphy, 
  Carswell}]{cwmc03}
Curran S.~J., Webb J.~K., Murphy M.~T.,  Carswell R.~F.
  2004{\natexlab{c}}, MNRAS, submitted (astro-ph/0311357)

\bibitem[{Curran {et~al.}(2004{\natexlab{d}})Curran, Webb, Murphy, 
  Kuno}]{cwmk03}
Curran S.~J., Webb J.~K., Murphy M.~T.,  Kuno N., 2004{\natexlab{d}}, in
  Millar T.~J., ed, ASP Conf. Ser., The Astrochemistry of External Galaxies, 
  San Francisco, in press (astro-ph/0310589)

\bibitem[{{Drinkwater} {et~al.}(1998){Drinkwater}, {Webb}, {Barrow} \& {Flambaum}}]{dwbf98}
{Drinkwater} M.~J., {Webb} J.~K., {Barrow} J.~D., {Flambaum} V.~V., 1998,
  MNRAS, 295, 457

\bibitem[{{Ellison} {et~al.}(2001){Ellison}, {Pettini}, {Steidel} \& {Shapley}}]{epss01}
{Ellison} S.~L., {Pettini} M., {Steidel} C.~C., {Shapley} A.~E., 2001,
  ApJ, 549, 770

\bibitem[{Ellison {et~al.}(2001)Ellison, Yan, Hook, Pettini, Wall, 
  Shaver}]{eyh+02}
Ellison S.~L., Yan L., Hook I.~M., Pettini M., Wall J.~V.,  Shaver P.
  2001, A\&A, 379, 393

\bibitem[{{Ge} {et~al.}(1997){Ge}, {Bechtold}, {Walker}, {Black}}]{gbwb97}
{Ge} J., {Bechtold} J., {Walker} C., {Black} J.~H., 1997, ApJ, 486, 727

\bibitem[{Kanekar  Chengalur(2003)}]{kc02}
Kanekar N.,  Chengalur J.~N., 2003, A\&A, 399, 857

\bibitem[{{Kovalev} {et~al.}(1999){Kovalev}, {Nizhelsky}, {Kovalev}, {Berlin},
  {Zhekanis}, {Mingaliev} \& {Bogdantsov}}]{knk+99}
{Kovalev} Y.~Y., {Nizhelsky} N.~A., {Kovalev} Y.~A., {Berlin} A.~B.,
  {Zhekanis} G.~V., {Mingaliev} M.~G., {Bogdantsov} A.~V., 1999, A\&AS,
  139, 545

\bibitem[{Kulkarni \& Fall(2002)}]{kf02}
Kulkarni V.~P.,  Fall S.~M., 2002, ApJ, 580, 732

\bibitem[{{Lanzetta} {et~al.}(1991){Lanzetta}, {Wolfe}, {Turnshek}, {Lu},
  {McMahon} \& {Hazard}}]{lwt+91}
{Lanzetta} K.~M., {Wolfe} A.~M., {Turnshek} D.~A., {Lu} L., {McMahon}
  R.~G., {Hazard} C., 1991, ApJS, 77, 1

\bibitem[{Ledoux {et~al.}(2003)Ledoux, Petitjean,  Srianand}]{lps03}
Ledoux C., Petitjean P.,  Srianand R., 2003, MNRAS, 346, 209

\bibitem[{{Liszt}(2002)}]{lis02}
{Liszt} H., 2002, A\&A, 389, 393

\bibitem[{{Liszt} \& {Lucas}(2000{\natexlab{a}})}]{ll00a}
{Liszt} H., {Lucas} R., 2000{\natexlab{a}}, A\&A, 355, 333

\bibitem[{{Liszt} \& {Lucas}(2000{\natexlab{b}})}]{ll00}
{Liszt} H.~S., {Lucas} R., 2000{\natexlab{b}}, A\&A, 355, 333

\bibitem[{Lovas(1992)}]{lov92}
Lovas F.~J., 1992, J. Phys. Chem. Ref. Data, 21, 181

\bibitem[{Lu {et~al.}(1999)Lu, Sargent,  Barlow}]{lsb99}
Lu L., Sargent W. L.~W.,  Barlow T.~A., 1999, in Carilli C.,
Radford S. Menton K., Langston G., eds, ASP Conf. Ser. Vol. 156,
Highly Redshifted Radio Lines, San Francisco, p. 132

\bibitem[{{Lu} {et~al.}(1996){Lu}, {Sargent}, {Barlow}, {Churchill} \& {Vogt}}]{lsb+96}
{Lu} L., {Sargent} W. L.~W., {Barlow} T.~A., {Churchill} C.~W., {Vogt}
  S.~S., 1996, ApJS, 107, 475

\bibitem[{Murphy {et~al.}(2001)Murphy, Webb, Flambaum, Drinkwater, Combes, 
  Wiklind}]{mwf+00}
Murphy M.~T., Webb J.~K., Flambaum V.~V., Drinkwater M.~J., Combes F., 
  Wiklind T., 2001, MNRAS, 327, 1244

\bibitem[{P\'{e}roux {et~al.}(2001)P\'{e}roux, Storrie-Lombardi, McMahon,
  Irwin,  Hook}]{psm+01}
P\'{e}roux C., Storrie-Lombardi L.~J., McMahon R.~G., Irwin M., Hook
  I.~M., 2001, AJ, 121, 1799

\bibitem[{{Petitjean} {et~al.}(2002){Petitjean}, {Srianand} \& {Ledoux}}]{psl02}
{Petitjean} P., {Srianand} R., {Ledoux} C., 2002, MNRAS, 332, 383

\bibitem[{Pickett {et~al.}(1998)Pickett, Poynter, Cohen, Delitsky, Pearson, 
  M\"{u}ller}]{ppc+98}
Pickett H.~M., Poynter R.~L., Cohen E.~A., Delitsky M.~L., Pearson J.~C.,
   M\"{u}ller H. S.~P., 1998, J. Quant. Spectrosc. Radiat. Transfer, 60, 883

\bibitem[{Prochaska {et~al.}(2003)Prochaska, Gawiser, Wolfe, Castro, 
  Djorgovski}]{pgw+03}
Prochaska J.~X., Gawiser E., Wolfe A.~M., Castro S.,  Djorgovski S.~G.
  2003, ApJ, 595, L9

\bibitem[{Rantakyr\"{o} {et~al.}(2004)Rantakyr\"{o}, Curran, Murphy,
  Staveley-Smith,  Webb}]{rcm+03}
Rantakyr\"{o} F., Curran S.~J., Murphy M.~T., Staveley-Smith L.,  Webb
  J.~K., 2004, MNRAS, in preparation

\bibitem[{{Rao} \& {Turnshek}(2000)}]{rt00}
{Rao} S.~M., {Turnshek} D.~A., 2000, ApJS, 130, 1

\bibitem[{Reimers {et~al.}(2003)Reimers, Baade, Quast,  Levshakov}]{rbql03}
Reimers D., Baade R., Quast R.,  Levshakov S.~A., 2003, A\&A, 410, 785

\bibitem[{Rohlfs \& Wilson(2000)}]{rw00}
Rohlfs K., Wilson T.~L., 2000, Tools of Radio Astronomy, 
  Springer-Verlag, Berlin

\bibitem[{{Savage} \& {Sembach}(1996)}]{ss96}
{Savage} B.~D., {Sembach} K.~R., 1996, Ann. Rev. Astr. Ap., 34, 279

\bibitem[{{Steppe} {et~al.}(1995){Steppe}, {Jeyakumar}, {Saikia} \& {Salter}}]{sjss95}
{Steppe} H., {Jeyakumar} S., {Saikia} D.~J., {Salter} C.~J., 1995, A\&AS,
  113, 409

\bibitem[{{Storrie-Lombardi} \& {Wolfe}(2000)}]{sw00}
{Storrie-Lombardi} L.~J., {Wolfe} A.~M., 2000, ApJ, 543, 552

\bibitem[{{Takahara} {et~al.}(1987){Takahara}, {Nakai}, {Briggs}, {Wolfe} \& {Liszt}}]{tnb+87}
{Takahara} F., {Nakai} N., {Briggs} F.~H., {Wolfe} A.~M., {Liszt} H.~S.
  1987, PASJ, 39, 933

\bibitem[{Takahara {et~al.}(1984)Takahara, Sofue, Inoue, Nakai, Tabara, 
  Kato}]{tsi+84}
Takahara F., Sofue Y., Inoue M., Nakai N., Tabara H.,  Kato T., 1984,
  PASJ, 36, 387

\bibitem[{{Ter\"{a}sranta} {et~al.}(1998){Ter\"{a}sranta}, {Tornikoski},
  {Mujunen}, {Karlamaa}, {Valtonen}, {Henelius}, {Urpo}, {Lainela}, {Pursimo},
  {Nilsson}, {Wiren}, {Laehteenmaeki}, {Korpi}, {Rekola}, {Heinaemaeki},
  {Hanski}, {Nurmi}, {Kokkonen}, {Keinaenen}, {Joutsamo}, {Oksanen},
  {Pietilae}, {Valtaoja}, {Valtonen} \& {Koenoenen}}]{ttm+98}
{Ter\"{a}sranta} et al., 1998, A\&AS, 132, 305

\bibitem[{Tornikoski {et~al.}(2000)Tornikoski, Lainela,  Valtaoja}]{tlv00}
Tornikoski M., Lainela M.,  Valtaoja E., 2000, AJ, 120, 2278

\bibitem[{{Tsuboi} \& {Nakai}(1991)}]{tn91}
{Tsuboi} M., {Nakai} N., 1991, PASJ, 43, L65

\bibitem[{{Turnshek} {et~al.}(2003){Turnshek}, {Rao}, {Ptak}, {Griffiths} \& {Monier}}]{trp+03}
{Turnshek} D.~A., {Rao} S.~M., {Ptak} A.~F., {Griffiths} R.~E., 
  {Monier} E.~M., 2003, ApJ, 590, 730

\bibitem[{{Turnshek} {et~al.}(1989){Turnshek}, {Wolfe}, {Lanzetta}, {Briggs},
  {Cohen}, {Foltz}, {Smith} \& {Wilkes}}]{twl+89}
{Turnshek} D.~A., {Wolfe} A.~M., {Lanzetta} K.~M., {Briggs} F.~H., {Cohen}
  R.~D., {Foltz} C.~B., {Smith} H.~E., {Wilkes} B.~J., 1989, ApJ, 344, 567

\bibitem[{White {et~al.}(1993)White, Kinney,  Becker}]{wkb93}
White R.~L., Kinney A.~L.,  Becker R.~H., 1993, ApJ, 407, 456

\bibitem[{{Wiklind} \& {Combes}(1994)}]{wc94}
{Wiklind} T., {Combes} F., 1994, A\&A, 286, L9

\bibitem[{{Wiklind} \& {Combes}(1995)}]{wc95}
{Wiklind} T., {Combes} F., 1995, A\&A, 299, 382

\bibitem[{{Wiklind} \& {Combes}(1996{\natexlab{a}})}]{wc96}
{Wiklind} T., {Combes} F., 1996{\natexlab{a}}, Nat, 379, 139

\bibitem[{{Wiklind} \& {Combes}(1996{\natexlab{b}})}]{wc96b}
{Wiklind} T., {Combes} F., 1996{\natexlab{b}}, A\&A, 315, 86

\bibitem[{{Wiklind} \& {Combes}(1996{\natexlab{c}})}]{wc96a}
{Wiklind} T., {Combes} F., 1996{\natexlab{c}}, in Shaver P. A., 
Science with Large Millimetre Arrays, Proceedings of the ESO-IRAM-NFRA-Onsala Workshop,
Springer-Verlag, Berlin, p. 86

\bibitem[{{Wiklind} \& {Combes}(1997)}]{wc97}
{Wiklind} T., {Combes} F., 1997, A\&A, 328, 48

\bibitem[{{Wiklind} \& {Combes}(1998)}]{wc98}
{Wiklind} T., {Combes} F., 1998, ApJ, 500, 129

\bibitem[{{Wiklind} \& {Combes}(1999)}]{wc99b}
{Wiklind} T., {Combes} F., 1999, in Carilli, C.,
Radford, S. Menton, K., Langston, G., eds, ASP Conf. Ser. Vol. 156,
Highly Redshifted Radio Lines, San Francisco, p. 202

\bibitem[{{Wolfe} {et~al.}(1995){Wolfe}, {Lanzetta}, {Foltz} \& {Chaffee}}]{wlfc95}
{Wolfe} A.~M., {Lanzetta} K.~M., {Foltz} C.~B., {Chaffee} F.~H., 1995,
  ApJ, 454, 698

\end{thebibliography}
\end{document}